\documentclass[12pt]{article}
\usepackage{graphics}

\catcode`@=11

\newcommand{\bq}{\begin{equation}}
\newcommand{\eq}{\end{equation}}

\newcommand{\bqa}{\begin{eqnarray}}
\newcommand{\eqa}{\end{eqnarray}}

\newcommand{\B}{\beta}
\newcommand{\G}{\gamma}
\newcommand{\Bbb}{B^{}_{\B\B}}
\newcommand{\Bbg}{B^{}_{\B\G}}
\newcommand{\Bgg}{B^{}_{\G\G}}
\newcommand{\Sg}{\sin\!3\G}
\newcommand{\dbe}{\partial_{\B}}
\newcommand{\dga}{\partial_{\G}}
\newcommand{\rnw}{\sqrt{r\over w}}
\newcommand{\nn}{\nonumber}

\begin{document}
 \title{\bf The Low-lying Quadrupole Collective \\
Excitations of Ru and Pd Isotopes\thanks{The work is supported in part by 
KBN contract No. 2 P03B 06813.}}
\author{K. Zaj\c ac, L. Pr\'ochniak and K. Pomorski\\
{\it Institute of Physics, The Maria Curie-Sk{\l}odowska University,}\\
{\it pl. M. Curie-Sk{\l}odowskiej 1, 20-031 Lublin, Poland}
\vspace{0.5cm}\\    
  S.G. Rohozi\'nski and J. Srebrny\\
{\it Department of Physics, Warsaw University,}\\
{\it Ho\.za 69, 00-681 Warsaw, Poland}}
\maketitle

\begin{abstract}
Quadrupole excitations of even-even Ru and Pd isotopes are described
within microscopic approach based on the general collective Bohr
model which includes the effect of coupling with the pairing vibrations.
The excitation energies and E2 transition probabilities observed in
$^{104-114}$Ru and $^{106-110}$Pd are reproduced in the frame of 
the calculation containing no free parameters. Particularly interesting
are $^{104}$Ru and $^{106-110}$Pd where good agreement with very rich 
information based on Coulomb excitation experiments is achieved.
\end{abstract}


\newpage
\section{Introduction}
\label{int}

The microscopic approach to the general collective Bohr hamiltonian
\cite{[Ku67],[Ro77]} was formulated many years ago in order to
investigate the coupling between nuclear rotations and surface
oscillations. This microscopic Bohr hamiltonian still offers a
consistent way to interpret nuclear collective modes directly referring
to single-particle degrees of freedom. Within this approach one can
construct realistic models of the collective quadrupole excitations in
different nuclei without introducing free parameters and fitting
procedures what is rather inevitable if calculations are made in the
frame of phenomenological descriptions (as, for instance, in
\cite{[Is80]} or \cite{[Tr96]}). This important feature allows to
investigate the structure of collective bands in the wide range of
even-even nuclei using no adjustable parameters except those fixed for
all nuclei: the single-particle potential parameters and the strength of
the residual pairing interaction.

However, the microscopic approach to nuclear collective excitations has
not been as useful as it could be because it failes in reproducing the
experimental level densities. As it is known since a long time
\cite{[Ro77]}, the energies of excited levels obtained within
microscopic Bohr model are in general larger than the measured ones. The
possible reason for such a disagreement found in Ref.~\cite{[Pi93]} was
the absence of an important collective degree of freedom connected with
the pairing vibrations \cite{[Be70]}. The strong influence of pairing
correlations on collective nuclear movements contributes to the coupling
between quadrupole and pairing vibrations. For that reason  the pairing
energy gap $\Delta$ should not be artifficially fixed at its equilibrium
point (found from the BCS equations) when studying nuclear collective
excitations. In fact, $\Delta$ should be treated as a collective
variable representing changes in the pairing field due to coherent
nucleon movements composing collective modes.

Recently we have developed in Ref.~\cite{[My]} an approximation based on
the general Bohr hamiltonian which includes the average effect of
pairing dynamics to the quadrupole excitations. Within this method
(described in Section~\ref{mod}) we have obtained the energies of excited
levels close to their experimental positions for the chains of
neutron-deficient Te, Xe, Ba, Ce and Nd isotopes. In the present work we
would like to show the results of calculations performed in the region
of neutron-rich deformed nuclei from the mass region of $A\approx 100$.
In Section~\ref{ru} we discuss the spectroscopic structure of $A=104-114$ Ru
isotopes which caused recently some controversions about the role of the
nonaxial $\G$ deformation. We have studied also  collective properties
of the $A=106-110$ Pd isotopes (Section~\ref{pd}) with the quadrupole band
structure disturbed by the influence of intruder states. For both groups
of nuclei we have obtained a significant succes in reproducing the
experimantal data. The results presented in this work confirm the
importance of the coupling with the pairing vibrations in proper
treatment of the nuclear collective quadrupole excitations.

\section{The model}
\label{mod}

In order to take into account the exact coupling between quadrupole
shape oscillations and pairing collective modes we should solve the full
"quadrupole + pairing" problem, what is rather troublesome because of 9
degrees of freedom to deal with: two intrinsic variables $\B$ and $\G$
parametrizing the shape of nuclear surface, three Euler angles for its
orientation in space (denoted in short as $\Omega$), two gap parameters
$\Delta^p$ and $\Delta^n$ for protons and neutrons and two corresponding
gauge angles $\phi^p,\phi^n$. Assuming that the coupling between
quadrupole and pairing variables is weak enough to neglect mixing terms
we can derive an approximate solution. The approximate collective
hamiltonian consists of two known terms (the operator $\hat{\cal H}_{\rm
int}$ mixing quadrupole and pairing variables will be neglected in
further calculations):
\bq
\label{hc}
\hat{\cal H}_{\rm CQP}  = \hat{\cal H}_{\rm CQ}(\B,\G,\Omega; \Delta
^p,\Delta^n) + \hat{\cal H}_{\rm CP}(\Delta^p,\Delta^n; \B,\G)
+ \hat{\cal H}_{\rm int}.
\eq
Here and in all following formulas the variables placed after a
semi-colon should be treated as parameters (they do not appear in
differential operators).
The first term $\hat{\cal H}_{\rm CQ}$ describes quadrupole 
oscillations and rotations of a nucleus and it takes the
form of the generalized Bohr hamiltonian \cite{[Ku67],[Ro77],[My]}
(in Ref.~\cite{[My]} it was denoted as $\hat{\cal H}_{\rm coll}$):
\bq
\label{hb}
\hat{\cal H}_{\rm CQ} = \hat{\cal T}_{\rm vib}(\B ,\G;\Delta^p,\Delta^n)
+ \hat{\cal T}_{\rm rot}(\B ,\G ,\Omega;\Delta^p,\Delta^n) + V_{\rm
coll}(\B,\G;\Delta^p,\Delta^n),
\eq
where $V_{\rm coll}$ is the collective potential. The kinetic
vibrational energy reads
\bqa
\label{vib}
\nn
\hat{\cal T}_{\rm vib}=-{{\hbar}^2\over{2\sqrt{wr}}}\bigg\{
{1\over \B^4}\bigg[ \dbe\bigg( \B^4\rnw \Bgg\dbe\bigg) -
\dbe \bigg(\B^3\rnw\Bbg\dga\bigg)\bigg]+ &&\\
+ {1\over \B\Sg}\bigg[ -\dga \bigg( \rnw\Sg\Bbg\dbe\bigg) +
{1\over\B}\dga \bigg(\rnw\Sg\Bbb\bigg)\dga\bigg]\bigg\}
\eqa
and the rotational energy is
\bq
\label{rot}
\hat{\cal T}_{\rm rot}={1\over 2}\sum_{k=1}^{3} \hat{I}^2_k/{\cal J}_k.
\eq
The intrinsic components of the total angular momentum are denoted as
$\hat{I}_k,\, (k=1,2,3)$, while $w$ and $r$ are the determinants of the
vibrational and rotational mass tensors. The mass parameters (or
vibrational inertial functions) $\Bbb$, $\Bbg$ and $\Bgg$ together with
moments of inertia ${\cal J}_k,\, (k=1,2,3)$ depend, in general, on
intrinsic variables $\B,\G$ and pairing gap values $\Delta^p,\Delta^n$.
All inertial functions are determined from a microscopic theory. We
apply the standard cranking method to evaluate the inertial functions
assuming that the nucleus is a system of nucleons moving in the deformed
mean field (Nilsson potential) which interact through monopole pairing
forces. According to the known formulas (given for example in
\cite{[My]}) the inertial functions can be expressed in terms of matrix
elements of derivatives of the nuclear hamiltonian on collective
variables, single-particle energies (depending on deformation point
$\B,\G$) and occupation probabilities of the single-particle levels
obtained by solving the BCS equations for given gap values
$\Delta^p,\Delta^n$. The collective potential is calculated within
Strutinsky macroscopic-microscopic method \cite{[Strut]}.

It should be stressed that in the minimum of the BCS energy with respect
of $\Delta$ one obtains the usual equilibrium gap values $\Delta^p_{eq},
\Delta^n_{eq}$ and the operator
\bq\label{b}
\hat{\cal H}_{\rm CQ}(\B,\G,\Omega;
\Delta^p=\Delta^p_{eq},\Delta^n=\Delta^n_{eq})
\eq
becomes exactly the same as the Bohr hamitonian known from Ref.
\cite{[Ku67]} or \cite{[Ro77]}.

For a given nucleus the second term in Eq. (\ref{hc}) describes collective
pairing vibrations of systems of $Z$ protons and $A-Z$ neutrons
\bq\label{p}
\hat{\cal H}_{\rm CP} = \hat{\cal H}^Z_{\rm CP} +
\hat{\cal H}^{A-Z}_{\rm CP}
\eq
and it can be expressed in the following form \cite{[Be70],[Go85]}:
\bq\label{hp}
 \hat{\cal H}^{\cal N}_{\rm CP}=-\frac{\hbar^2}
{2\sqrt{g(\Delta)}}\frac{\partial}
{\partial\Delta}\frac{\sqrt{g(\Delta)}}
{B_{\Delta\Delta}(\Delta)}
\frac{\partial}{\partial\Delta} + V_{\rm pair}(\Delta),
\eq
where ${\cal N}=Z$, $\Delta=\Delta^p$ for protons and, respectively,
${\cal N}=A-Z$, $\Delta=\Delta^n$ for neutrons. The functions appearing
in the hamiltonian (\ref{hp}), namely the pairing mass parameter
$B_{\Delta\Delta}(\Delta)$, the determinant of the metric tensor 
$g(\Delta)$ and the collective pairing potential $V_{\rm pair}(\Delta)$ 
are determined
microscopically at each deformation point $\B,\G$ according to the
formulas given in Ref. \cite {[Go85]}. The approximate projection of the
BCS wave function on correct particle number \cite{[My],[Go86]} is
applied within the calculation.


Solving the eigenproblem of the collective pairing hamiltonian
(\ref{hp}) one can find the pairing vibrational ground-state wave
function $\Psi ^{\cal N}_0$ and the ground-energy $E^{\cal N}_0$ at each
deformation point. The most probable value of the energy gap $\Delta_{vib}$
corresponds to the maximum of the probability  of finding a given gap
value in the collective pairing ground-state (namely the maximum of the
function $g(\Delta )|\Psi ^{\cal N}_{0}(\Delta )|^2$). As it is shown in
Fig.~1 the $\Delta_{vib}$ is shifted towards smaller gaps from the
equilibrium point $\Delta _{eq}$ determined by the minimum of $V_{\rm
pair}$ (or by the BCS formalism). Such a behaviour of the pairing ground
state function $\Psi ^{\cal N}_0$ is due to the rapid increase of
pairing mass parameter $B_{\Delta \Delta}$ and it appears in most cases.
In general the ratio of $\Delta _{vib}$ to $\Delta _{eq}$ is of about
$0.7$.

However, the exact diagonalization of the collective hamiltonian
(\ref{hc}) still remains difficult because of the dimension of the
needed basis. But as far as only low-lying nuclear excitatons are taken
into account one can consider the coupling between quadrupole and
pairing collective degrees of freedom diagonalizing the hamiltonian
(\ref{hb}) just in the point corresponding to the most probable gaps
$\Delta^p_{vib}$, $\Delta^n_{vib}$, i.e.
\bq\label{h}
\hat{\cal H}_{\rm CQP} \approx \hat{\cal H}_{\rm Q}(\B,\G,\Omega; \Delta^p
=\Delta^p_{vib},\Delta^n=\Delta^n_{vib}).
\eq
It means that all collective functions appearing in Eq. (\ref{h}) are
calculated using the most probable pairing gap values for protons and
for neutrons instead the equilibrium ones as it used to be when the
expression (\ref{b}) had been diagonalized. The collective potential
also depends on the most probable pairing gaps and, in addition, it is
slightly corrected by the pairing vibrational ground-state energies
$E^Z_0$ and $E^{A-Z}_0$ coming from the collective pairing term
(\ref{hp}). In order to solve the eigenproblem of the hamiltonian
$\hat{\cal H}_{\rm CQP}$ we calculate its matrix elements using the 
functions  $\tilde{\Phi}^{IM}_{j}(\B,\G,\Omega)$
where $I$ means the angular momentum, $M=-I,...,I$ while $j$ means the
set of additional quantum numbers. The basis 
$\tilde{\Phi}^{IM}_{j}(\B,\G,\Omega)$ was obtained 
\cite{[My]} following the approach of Libert and Quentin \cite{[Li82]}.

Other collective properties are also characterized by matrix elements of
appropriate operators evaluated in the basis mentioned above. All
collective operators (as the electric quadrupole moments) 
are determined microscopically using cranking formulas and
the most probable pairing gaps $\Delta^p_{vib},\Delta^n_{vib}$ \cite{[My]}.

The approximation described above is rather crude but, as we would like
to exemplify in next sections, it includes the main effect (at least on
average) of the coupling with the pairing vibrational mode. This
procedure improves significantly the accuracy in reproducing the
experimental data but it introduces {\it no additional parameters} into
the model. Our calculations were done using the standard Nilsson single
particle potential with the shell dependent parametrization found in
Ref. \cite{[Se86]}. For the pairing strength we applied the same
estimates as used in Ref. \cite {[My]}: $G= g_{0}\/{\cal N}^{2/3}$,
where ${\cal N}=Z$ or ${\cal N}=A-Z$ for protons or neutrons
respectively and $g_{0}=0.26\,\hbar\omega_{0}$.

\section{Ru isotopes}
\label{ru}

Recent investigations of the collective properties of even-even
neutron-rich Ru isotopes done in Ref. \cite {[Sh94]} and Ref.
\cite{[Lu95]} caused a discussion about the role of the $\G$ deformation
in this region. These nuclei appear to be generally triaxial:
both, predicted \cite{[Sk97]} and experimentally deduced by fitting to
the rigid triaxial rotor model \cite{[Sh94]} the equilibrium $\G$ values
are close to $20^{\circ}$. On the other hand, there were made some
observations \cite {[Lu95]} suggesting that Ru isotopes are rather
$\G$-soft with prolate equilibrium shapes. In consequence different
(even opposite) phenomenological approaches as the rigid triaxial rotor
model (asymmetric rotor model) \cite{[Tr96],[Sh94],[Ei70]},
the rotation-vibration model \cite{[Tr96],[Lu95],[Ei70]} 
and more general (but still based on the geometrical approach)
collective model adopted in Ref.~\cite {[Tr96]} were applied
in order to interpret collective bands in Ru isotopes, each with a
comparable success. It seems that the reason of such a situation may be
connected with some essential difficulties in determining the
equilibrium shapes. The collective
potentials calculated within the model described in Section~\ref{mod} appear
generally triaxial (see Fig.~2), what is in agreement with the
expectations \cite{[Sh94],[Sk97]}. However, the bottoms of the potential
energy surfaces are very flat what can lead to the considerable
uncertainty in location of the minima. Anyway, a $\G$-softness of the 
nuclei in question seems to be beyond all doubts.


Diagonalizing the hamiltonian (\ref{h}) with such collective potentials
we have obtained the excitation energies and the corresponding wave
functions of $^{104-114}$Ru (dependent on $\B$ and, rather smoothly, on 
$\G$ deformation). The calculated energies of positive parity
states of $^{104}$Ru presented in Fig.~3 (see the part marked as "new")
agree very well with the measured ones. Small discrepancies appearing
for the higher $0^+$ states may be connected with the absence of mixing
terms in the model hamiltonian (\ref{hc}) or with the restrictions
imposed on the basis (see Section~\ref{mod}). For the $8^+_2$ level the 
difference between theoretical and experimental \cite{[St84],[Sr84]} energy 
is rather due to the observed mixture of two-particle mode.
For comparison we present also (see the part of Fig.~3 marked as "old") 
the excitation energies obtained with the original Bohr hamiltonian 
(expression \ref{b} in Section~\ref{mod}). As one can learn from Fig.~3  
the improvement in reproducing the experimental data caused by 
coupling with the pairing vibrations is really significant.


To get a real proof of the proper identification of obtained wave 
functions we have investigated their
electromagnetic properties. The experimental evidence gathered
by Coulomb excitations in Ref. \cite{[St84],[Sr84]} for $^{104}$Ru 
is fairly rich  so we can verify our
results in details. Calculated diagonal matrix elements of the
quadrupole electric operator in the excited states of this nucleus are
compared with the observed ones in Fig.~4. Theoretical values
follow the experimental data, the agreement is good enough
to reproduce almost exactly the the changement of the quadrupole moment of
$^{104}$Ru along the bands.


In Fig.~5 we present theoretical and experimental reduced
E2 transition probabilities. The transitional probabilities 
observed in the ground-state and $\G$ bands were measured by 
J.~Stachel et al. \cite{[St84]} while E2 transitions involving the $0^+_2$
band were investigated by J.~Srebrny et al. \cite{[Sr84]}. The results 
of calculations show that the wave functions obtained in our 
approximation really represent low-lying excited states of $^{104}$Ru.
We have obtained too small values of the probabilities of E2 transitions
between members of the ground-state band but they are still close to the
data. Besides, for the $\G$ and $0^+_2$ bands we have reached a very good
agreement with the observed E2 transition probabilities. Even
some  weak probabilities of the band-to-band transitions obtained in 
this work are
nicely confirmed by the experiment.


We have obtained a similar agreement with the experimental data
\cite{[Sh94],[Lu95],[Si98],[data]} for excited states in $^{106,108}$Ru
(Fig.~6)
and $^{110,112}$Ru (Fig.~7) isotopes. There are some discrepancies, for
example the calculated $6^+_1$ level descends a little below the
experimental point disturbing the smooth experimental dependence of the
energy on the angular momentum. It looks like a sudden change in the
structure of the ground-state band energies which is not seen in the
experiment. Perhaps it is due to the lack of components of higer
multipolarities in our model. Another inaccuracy appears in $^{112}$Ru,
where the moment of inertia in the $\G$ band is too large with respect
to observed one. In spite of those relatively small discrepancies we
have got a realistic description of the collective band structure in all
considered Ru isotopes, even for such a very neutron reach nucleus as
$^{114}$Ru (Fig.~8).


As was mentioned before, the lowering of quadrupole transitions between
the neighbouring members of a ground-state band seems to be a general
feature of our model in the region of neutron-deficient nuclei
\cite{[My]}. This tendency occurs also in $^{106-114}$Ru isotopes: few
available experimental data points are situated (see Fig.~9) above the
theoretical points. The experimental evidence of the reduced transition
probabilities in heavy Ru isotopes is rather poor but we suppose that
our predictions should be as realistic as for $^{104}$Ru, especially the
results obtained for $\G$ bands. We did not find any data concerning
quadrupole moments in the excited $^{106-114}$Ru states but we would
like to present in Fig.~10 the predicted (and rather characteristic for
all considered Ru isotopes) dependence of the diagonal E2 diagonal matrix
elements on the angular momentum.


In the last years there were made rather intensive theoretical
investigations of $^{108-112}$Ru within different phenomenological
models, e.g. the triaxial rotor model \cite{[Sh94],[Tr96]}, 
the rotation --vibration model \cite{[Lu95],[Tr96]} and
the generalized collective model (GCM) \cite{[Tr96]}. All
those models have numerous parameters fitted to the data (for instance 8
parameters for each nucleus in the GCM). Thus the direct comparison 
of earlier results with ours which are obtained without adjustable
parameters does not seem to be very useful.
Nevertheless, within approach presented here the calculated
energies of excited levels in $^{108-112}$Ru are situated almost as
close to their experimental positions as in Ref. \cite{[Tr96]}.

The accuracy in reproducing the observed band structure suggests almost
pure quadrupole nature of low-lying states in the $Z=44, A=104-112$
nuclei. The further suggestion is that their shapes are in general
triaxial but soft or very $\G$-soft.
The coupling between quadrupole and pairing collective degrees
of freedom plays an essential role in description of the neutron-rich Ru
isotopes but, on the other hand, the shape and pairing vibrations can be
treated separately to some extent.

\section{Pd isotopes}
\label{pd}

As Ru isotopes were treated as triaxial rotors, the Pd isotopes used to
be interpreted in terms of vibrational modes. But the careful Coulomb 
excitation experimental investigations (e.g.
\cite{[Sv95],[Sv89],[Ha86]}) have changed this
picture and showed that Pd isotopes exhibit a rather complicated
structure. The IBA-2 model \cite{[Is80],[Sv89]} has been applied in
order to interpret the band-like spectroscopic structure and to
calculate quadrupole electric transitions (see Fig.~13). But it should be
mentioned that the IBA-2 model made use of 15 (and even more) free
parameters while the simpler IBA-1 version was unable to describe the Pd
nuclei because of a competition between SU(5) and O(6) boson limits
\cite{[Sv89],[St82]}.

Some complications in the band structure of Pd isotopes arise because of
the intruder states \cite{[Sv89],[We94]} and their interaction 
with "normal" single-particle states. But according to the
experimental suggestions the transitional $Z=46$ nuclei should be
interpreted using collective potentials soft in both $\B$ and $\G$
deformations. The collective Bohr hamiltonian (see Section~\ref{mod})
automatically takes into account the rotation-vibration coupling and it
seems to be an appropriate approach. Of course, also here the Bohr
hamiltonian should be generalized in order to describe the influence of
pairing vibrations, at least on average.


The $^{106-110}$Pd have been described in the frame of exactly the same
model which was referred in Section~\ref{mod} and applied to Ru isotopes in
Section~\ref{ru}. As a result we have obtained the low-lying positive parity
states, their energies, quadrupole moments and transitions for Pd isotopes. 
Collective potentials obtained for $^{106}$Pd and$^{108}$Pd are shown in
Fig.~11. The potential surfaces are very shallow and it is
almost imposible to determine exactly $\B$ and $\G$ equilibrium values.
The uncertainty of nuclear equilibrium shapes as well as the structure 
of wave functions corresponding to low excitation energies testify the 
softness of Pd isotopes in both, $\B$ and $\G$ deformations.


The theoretical energy levels are compared to the four quasi-rotational
bands observed in $^{106}$Pd, $^{108}$Pd (Fig.~12) and $^{110}$Pd
(Fig.~13). It is visible that ground state bands and $\G$ bands with
their characteristic staggering are reproduced very well. For remaining
bands the agreement is not so good. But the $0^+_2$ bands observed in
$^{108}$Pd and $^{110}$Pd have different features than it is expected for
collective states. According to Ref. \cite{[Sv89],[We94]} the large
deformations estimated experimentally for these bands can be due to the
particle-hole configurations connected with the presence of intruder
states. On the other hand, there are strong experimental indications
\cite{[Sv89]} that the nature $0^+_2$ band in $^{106}$Pd and the system
of levels built on the $0^+_3$ state in $^{110}$Pd are very similar.
Following this suggestion we decided to compare the theoretical $0^+_2$
, $2^+_3$, $4^+_3$ and $6^+_3$ levels to the members of the
experimentally found $0^+_3$ band. Such a new theoretical $0^+_3$ band
matches the observed one as is pictured in Fig.~13.


The rearrangement could be justified by investigations of the
electromagnetic properties of considered states. The reduced E2
transition probabilities between members of redefined theoretical
$0^+_3$ band agree much better with the experimental data (see Fig.~14)
than probabilities calculated in the ground state band (which tend to be
too small as in the other nuclei \cite{[My]}). However, the observed
diagonal matrix elements of E2 operator in all states represented in
Fig.~15 are reproduced quite well within our approximation.


Quadrupole electric transitions and diagonal matrix E2 elements for
$^{106,108}$Pd are compared to the experimental data in Fig.~16--18. 
The agreement reached in this work is not worse than
in the IBA-2 calculations using fitting procedures 
\cite{[Is80],[Sv89],[St82]}. In fact, we could find that our results 
obtained with no free parameters are even more reliable.


The structure of Pd isotopes is more complicated than the Ru ones:
several different degrees of freedom
are expected in the low-lying positive parity states in the Pd nuclei
including non-collective modes as the proton two- particle- two-
hole excitations. 
Nevertheless, we were able to interpret nuclear states hardly described
within  geometrical pictures of the vibrator or asymmetric rotor or
within IBA-1 model \cite{[St82]}. Our results support the picture of
the Pd isotopes as $\B$- and $\G$-soft quadrupole deformed nuclei 
easily adopting triaxial shapes. We found that the coupling 
betweeen quadrupole and pairing vibrations in Pd isotopes appears 
very important and plays the similar role as in Ru nuclei.

\section{Summary and Conclusions}

The general collective Bohr hamiltonian including the average effect of
coupling with the pairing vibrations was applied to neutron-rich Ru and
Pd isotopes. The collective properties of low-lying excited states were
interpreted within model containing no adjustable parameters. The
resulting excitation energies agree well with the measured values and,
moreover, a resonable agreement with the spectroscopic data was reached
in description of their electromagnetic properties. We should mention
that our approximation works even better for $A\approx 100$ neutron rich
nuclei than in the region of transitional neutron-deficient Te, Xe or Ba
isotopes \cite{[My]}.

Some discrepancies found in comparison with the experimental data are
partly due to the non-collective nature of some excited states.
Disagreements were observed especially for the configurations involving
intruder orbitals (the $0^+_2,0^+_3$ bands in $^{108,110}$Pd). Some
discrepancies could be also connected with the restrictive assumptions made
when the model hamiltonian was  formulated (see Section~\ref{mod}). The
coupling between quadrupole and pairing vibrations was approximated by
shifting of the pairing gap to its most probable value. This simplified
way may be too drastic to get appropriate values of reduced E2
transition probabilities. Especially too low values of the calculated
transition probabilities between members of ground state band are
supposed to be connected with the approximate treatment of the coupling
between pairing and quadrupole vibrations.

Nevertheless, the results of our calculations are close to the
experimental data. This agreement confirm the importance of the
collective pairing mode in the theoretical description of even-even
nuclei. We expect that the model could be improved by more careful
treatment of terms mixing the intrinsic variables $\B$ and $\G$ with
pairing gaps in the collective hamiltonian.

\newpage

{\bf Figure captions:}\\
\vspace{0.5cm}

{\bf Fig.~1} The pairing vibration mass parameter ($B_{\Delta\Delta}$),
and potential ($V_{\rm pair}$), and the ground-state function ($\Psi ^N_0$)
in dependence on the pairing energy gap $\Delta$ for the system of 
$60$ neutrons at the deformation point $\B = 0.2,\, \G = 20^{\circ}$. 
The equilibrium value of the energy gap is $\Delta_{eq} 
\approx 0.14\hbar\omega_0$, the most
probable one is $\Delta_{vib}\approx 0.09\hbar\omega_0$. \\

{\bf Fig.~2} The collective potential (corrected by the zero-point 
pairing vibrations) calculated for $^{106-112}$Ru isotopes. The energy
distance between neighbouring lines is of 1 Mev.

{\bf Fig.~3} The experimental \cite{[Sr84],[St84]} and the theoretical 
(connected by straight lines) excitated levels in $^{104}$Ru versus angular
momentum $J^{\pi}$. The theoretical values were calculated including
the effect of coupling with the pairing vibrations ("new") and 
without this coupling, i.e. within usual microscopic Bohr model ("old").\\

{\bf Fig.~4} The experimental \cite{[Sr84]} and the calculated reduced
diagonal matrix elements
of the quadrupole electric operator in excited states of $^{104}$Ru.
The theoretical points are connected by straight lines.\\

{\bf Fig.~5} Theoretical reduced E2 transition probabilities 
(white points connected by straight lines) in comparison to the values 
measured (black points) for the ground-state and $\G$ band \cite{[St84]}
and for $0^+_2$ band \cite{[Sr84]} in $^{104}$Ru. $J^{\pi}$ means
the angular momentum of an initial state.\\

{\bf Fig.~6} The experimental \cite{[Sh94],[data]} and the 
theoretical (marked with straight lines) energies of excited levels 
in $^{106}$Ru and $^{108}$Ru versus angular momentum $J^{\pi}$.\\

{\bf Fig.~7} The same as in Fig.~5 but for $^{110}$Ru and $^{112}$Ru.
Experimental data taken also from \cite{[Si98]}.\\

{\bf Fig.~8} The experimental \cite{[Sh94]} and the calculated
(marked with straight lines) energies of excited levels in $^{114}$Ru.\\

{\bf Fig.~9} Theoretical reduced probabilities of E2 transitions between
states of ground band (squares) and of $\G$ band (triangles) in
$^{106-114}$Ru. Few experimental data known for $2^+_1\rightarrow 0^+_1$
and $4^+_1\rightarrow 2^+_1$ transitions \cite{[data]} are marked with
the full squares.\\

{\bf Fig.~10} The predicted diagonal matrix elements of the quadrupole electric
operator in the states of the ground band (squares) and of $\G$ band
 (triangles) in Ru isotopes.\\

{\bf Fig.~11} The collective potential (including the zero-point 
pairing vibrations) calculated for $^{106-108}$Pd. The distance 
between neighbouring lines is of 1 Mev.\\

{\bf Fig.~12} The experimental \cite{[Sv95],[Sv89]} and the calculated excited
levels in $^{106,108}$Pd versus angular momentum $J^{\pi}$. 
The theoretical points are connected by straight lines.\\

{\bf Fig.~13} The experimental \cite{[Sv89],[Ha86]} and the calculated excited
levels in $^{110}$Pd versus angular momentum $J^{\pi}$. 
The theoretical points are connected by straight lines.
The calculated levels built on the second $0^+$ state are 
interpreted as the members of the $0^+_3$ band (see Section~\ref{ru}).\\

{\bf Fig.~14} The experimental \cite{[Sv89],[Ha86]} and the calculated
reduced probabilities of E2 in-band transitions in $^{110}$Pd. 
Theoretical points obtained in this work are connected with
straight lines. For comparison there are showed
results \cite{[Sv89]} of the IBA-2 calculation (points marked by dotted
lines).\\

{\bf Fig.~15} The experimental \cite{[Sv89],[Ha86]} and the calculated diagonal matrix
elements of the quadrupole electric operator in the ground, $\G$ and 
$0^+_3$ bands of $^{110}$Pd. The theoretical points are connected 
with straight lines.\\

{\bf Fig.~16} The experimental \cite{[Sv95],[Sv89]} and the calculated reduced
probabilities of E2 in-band transitions in $^{106,108}$Pd. The 
theoretical points are marked by straight lines.\\

{\bf Fig.~17} The experimental \cite{[Sv95],[Sv89]} and the calculated diagonal matrix
elements of the quadrupole electric operator in the ground and $\G$ 
bands of $^{106,108}$Pd. The theoretical points are marked by straight 
lines.  \\

{\bf Fig.~18} The experimental \cite{[Sv95],[Sv89],[Ha86]} and the calculated reduced
probabilities of E2 transitions from the $J^{\pi}$ state of the $\G$ 
band in $^{106-110}$Pd to the state of the same angular momentum in 
the corresponding ground band. The theoretical points are marked with 
straight lines.

\newpage

\framebox{Fig.~1}

\vspace{2cm}
{\includegraphics*{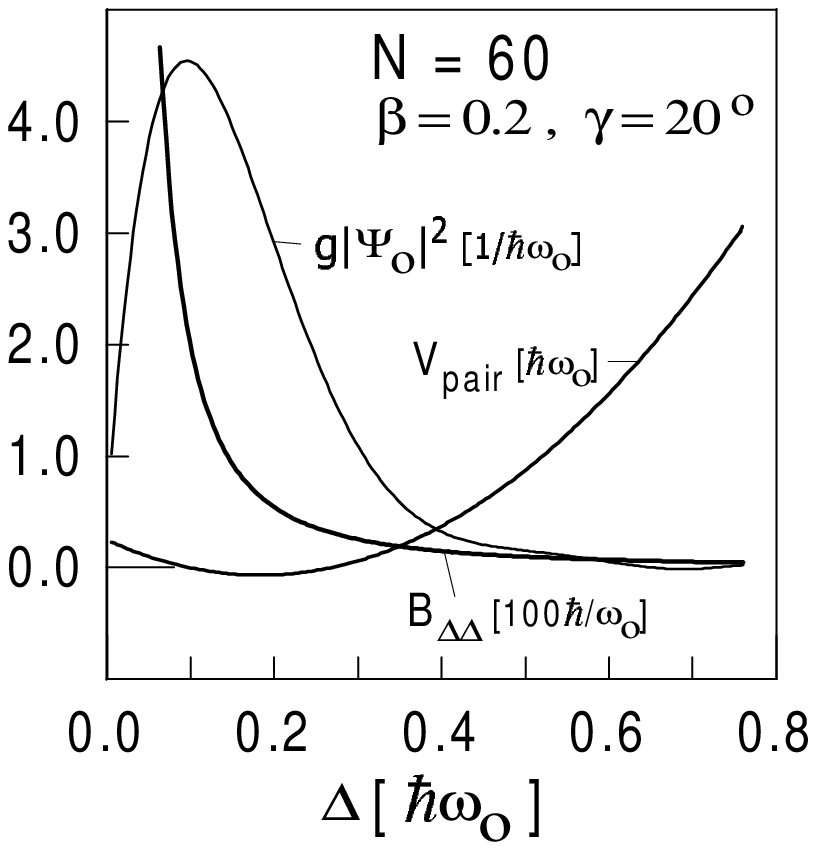}}

\newpage

\framebox{Fig.~2}

\vspace{2cm}
{\includegraphics*{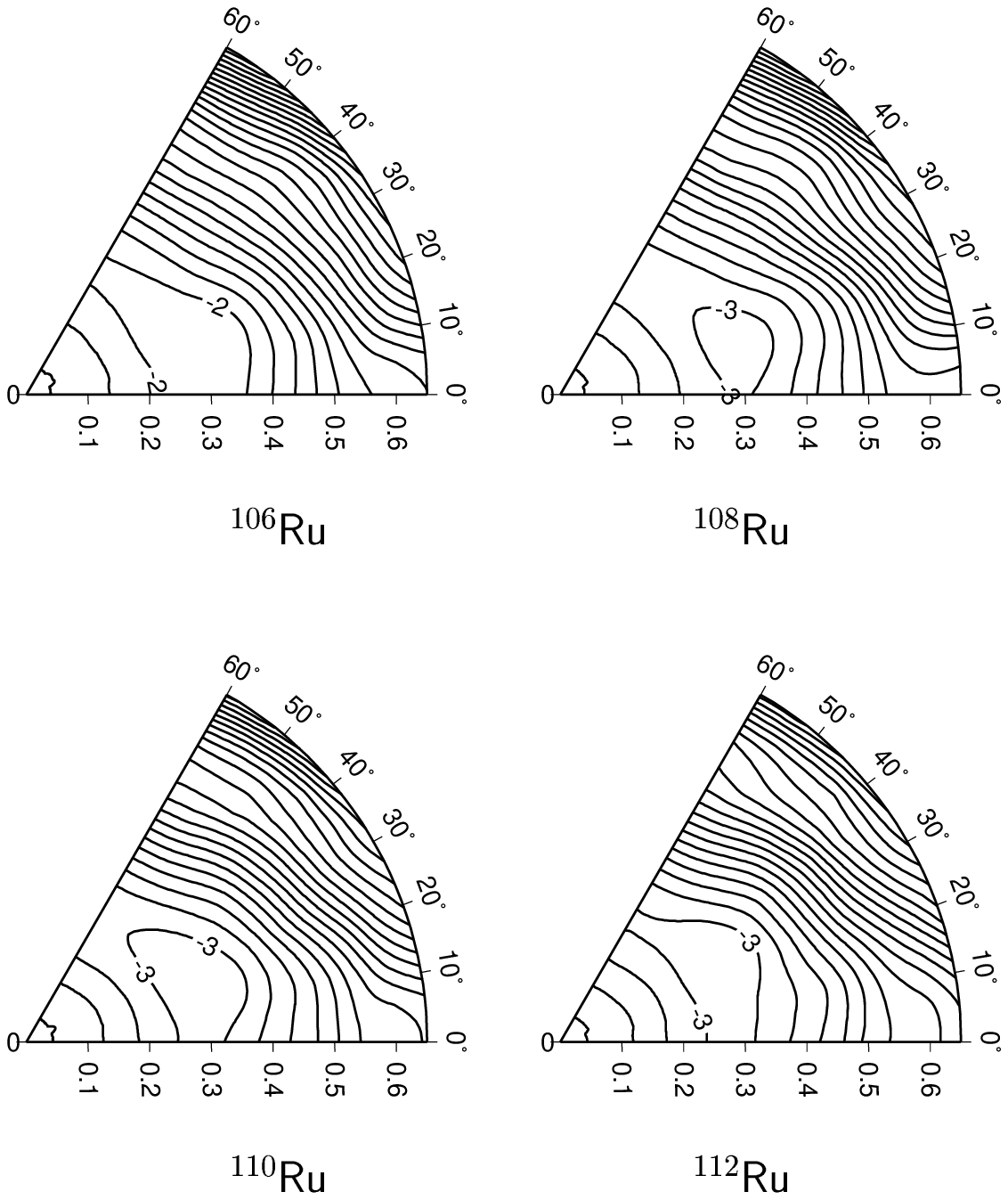}}
\newpage

\framebox{Fig.~3}

\vspace{2cm}
{\includegraphics*{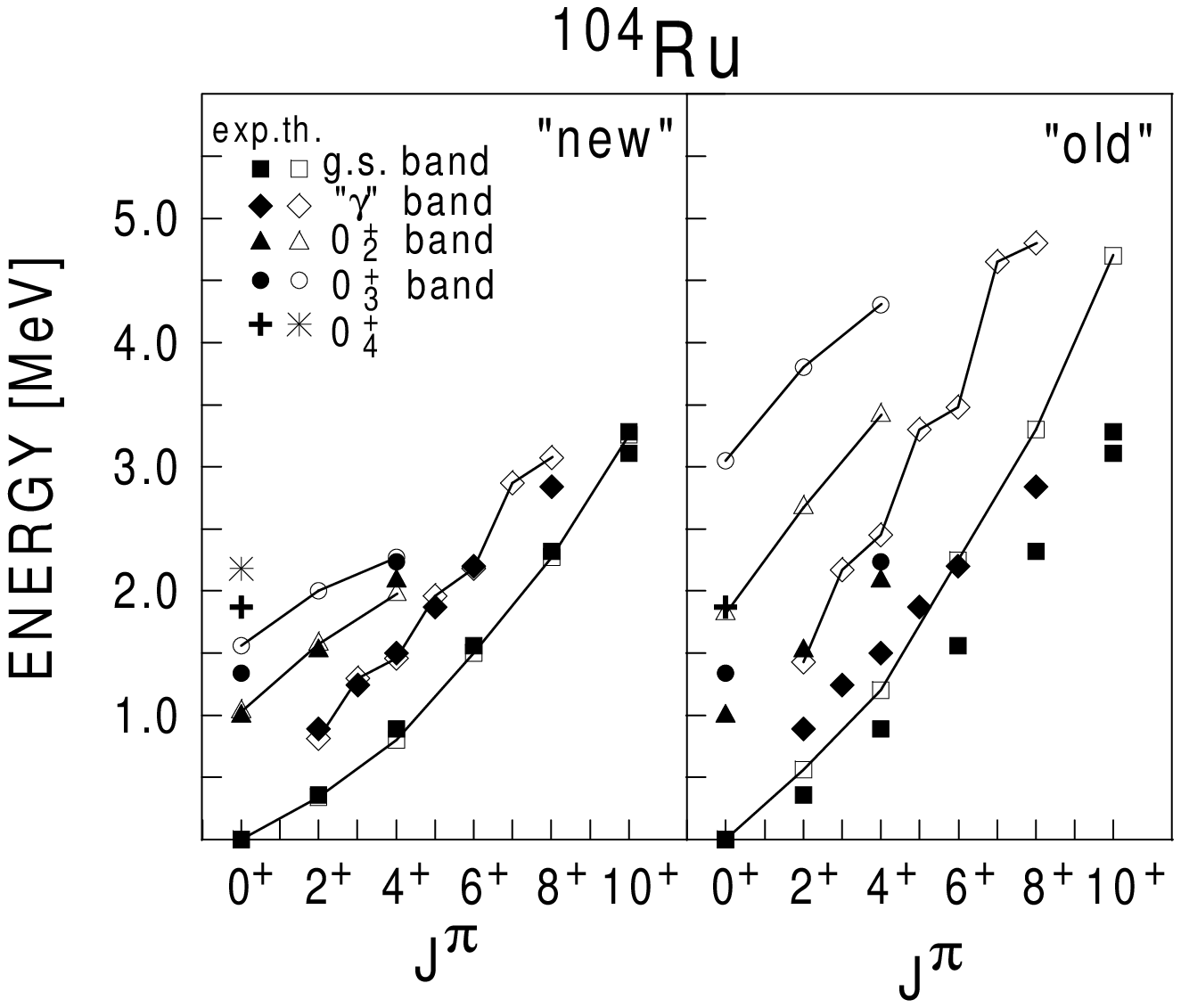}}
\newpage

\framebox{Fig.~4}

\vspace{2cm}
{\includegraphics*{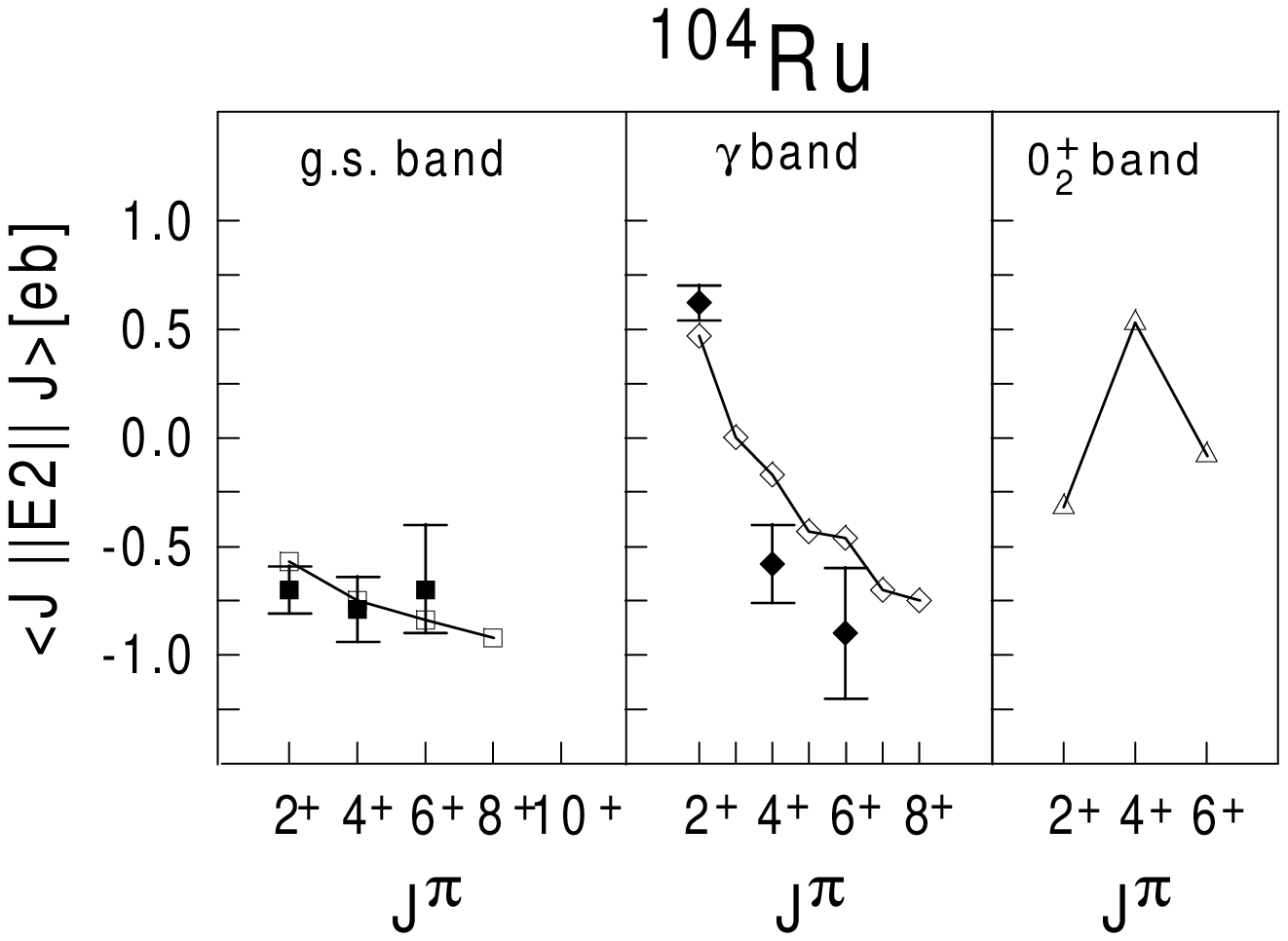}}
\newpage

\framebox{Fig.~5}

\vspace{2cm}
{\includegraphics*{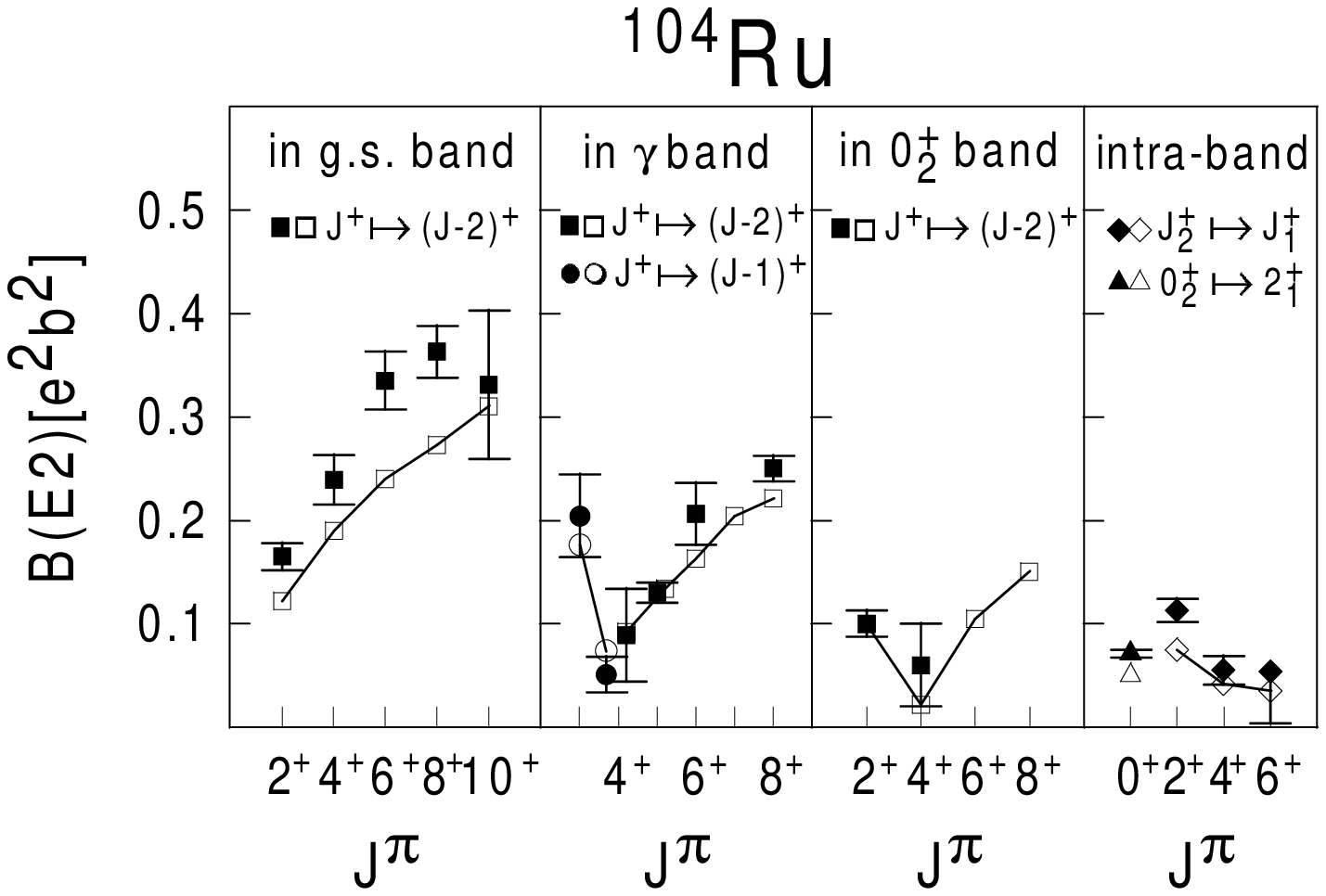}}
\newpage

\framebox{Fig.~6}

\vspace{2cm}
{\includegraphics*{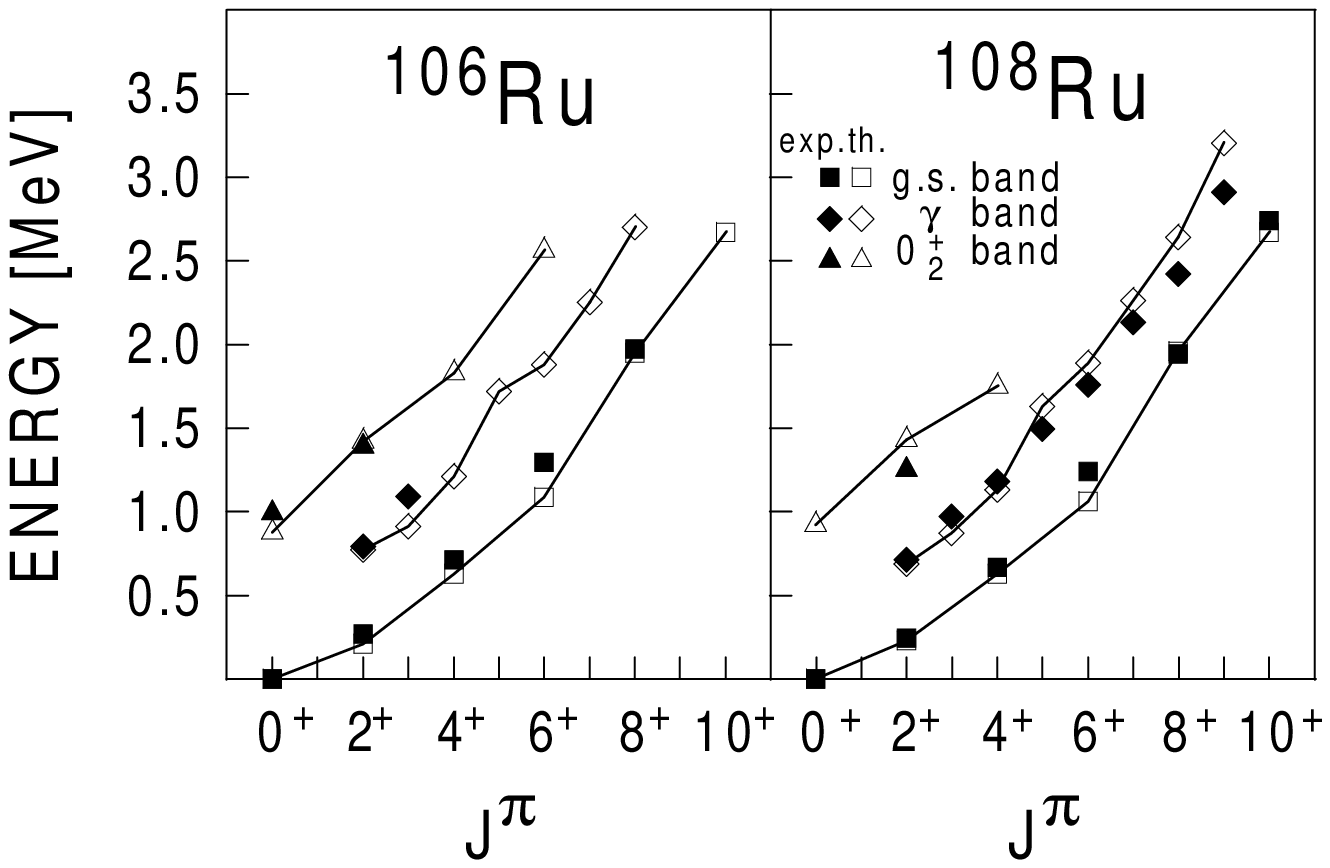}}

\newpage

\framebox{Fig.~7}

\vspace{2cm}
{\includegraphics*{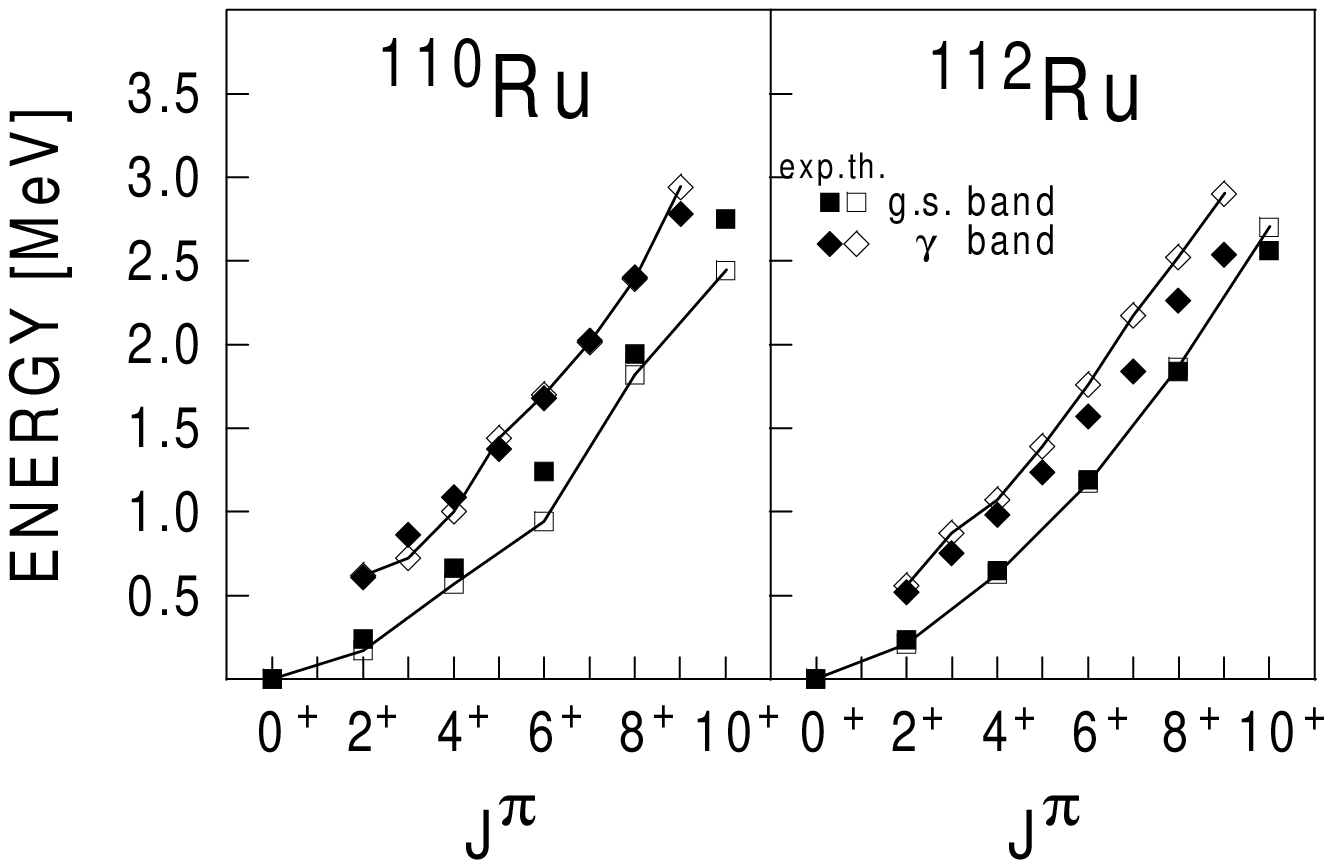}}

\newpage

\framebox{Fig.~8}

\vspace{2cm}
{\includegraphics*{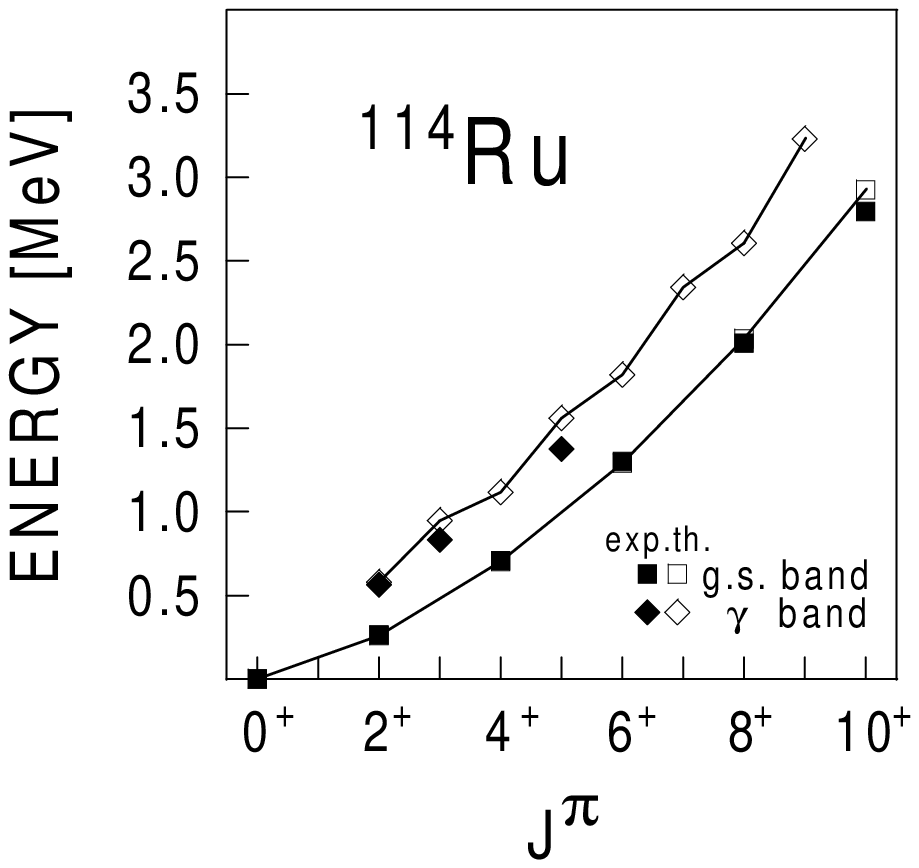}}
\newpage

\framebox{Fig.~9}

\vspace{2cm}
{\includegraphics*{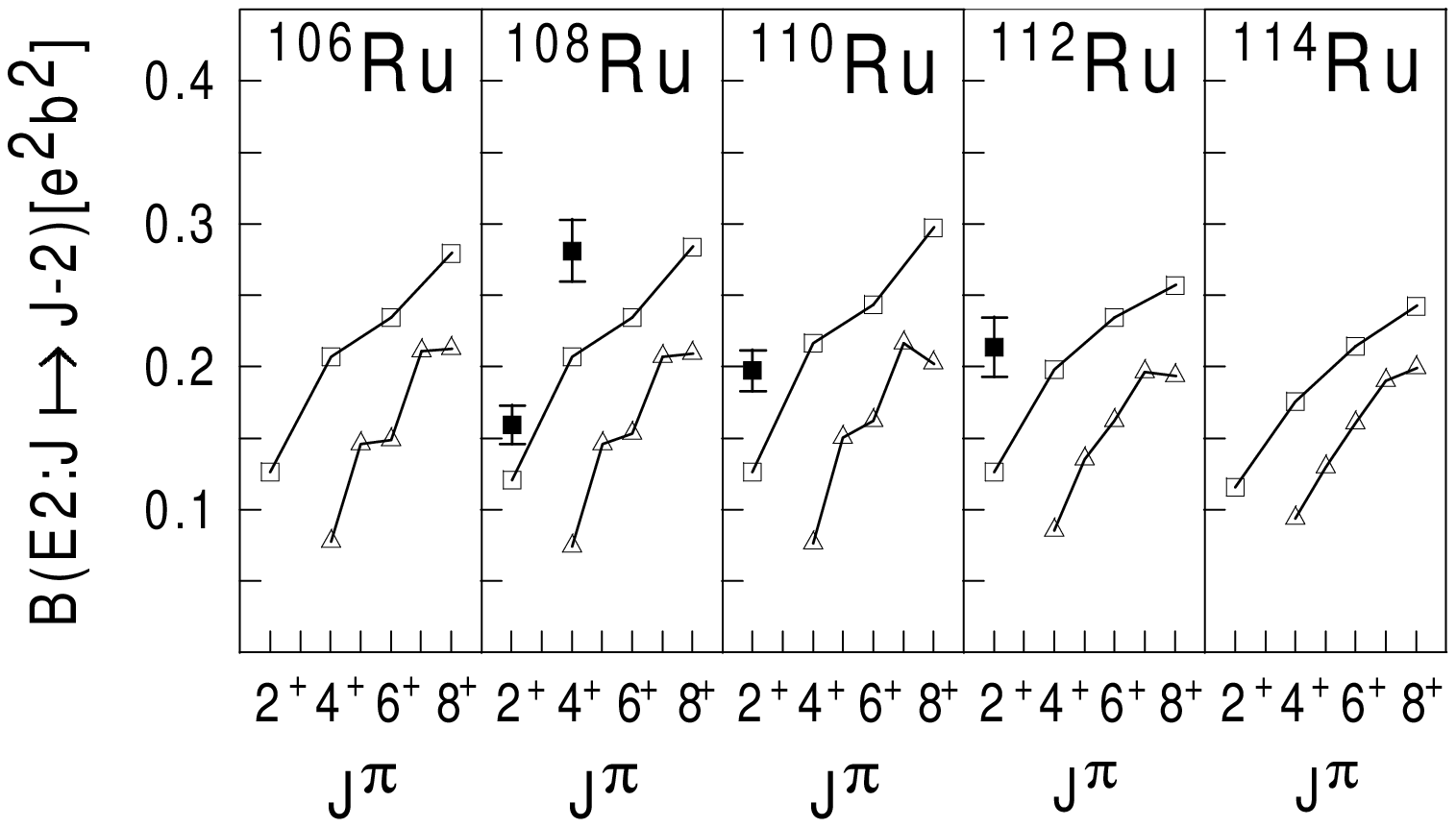}}
\newpage

\framebox{Fig.~10}

\vspace{2cm}
{\includegraphics*{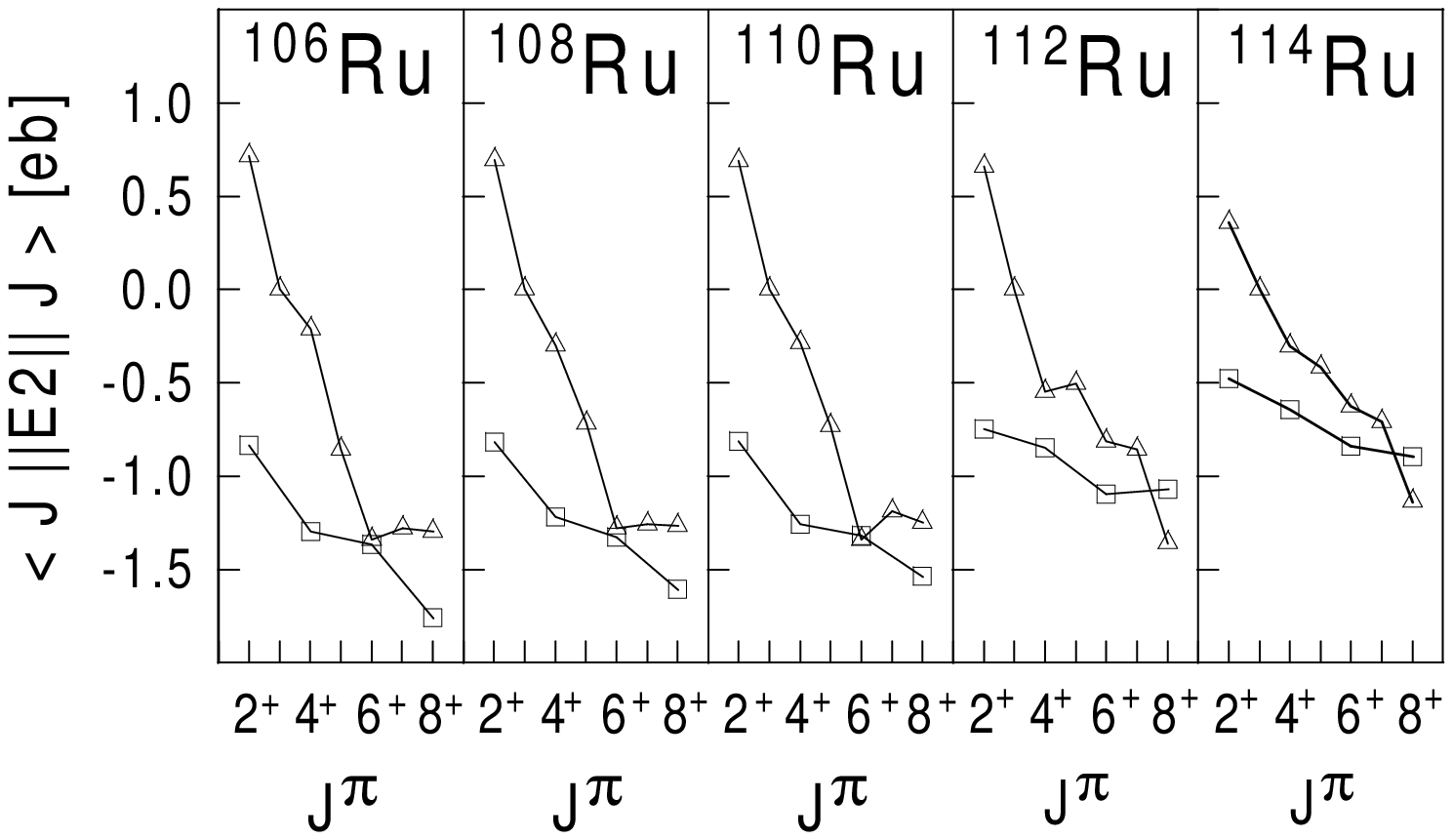}}
\newpage

\framebox{Fig.~11}

\vspace{2cm}
{\includegraphics*{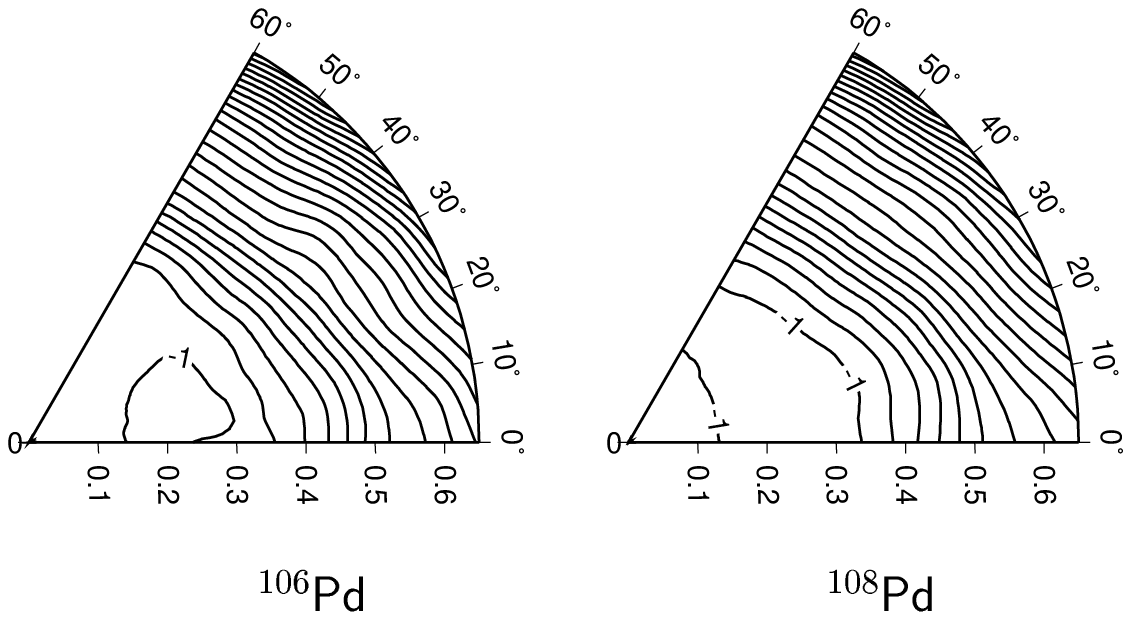}}
\newpage

\framebox{Fig.~12}

\vspace{2cm}
{\includegraphics*{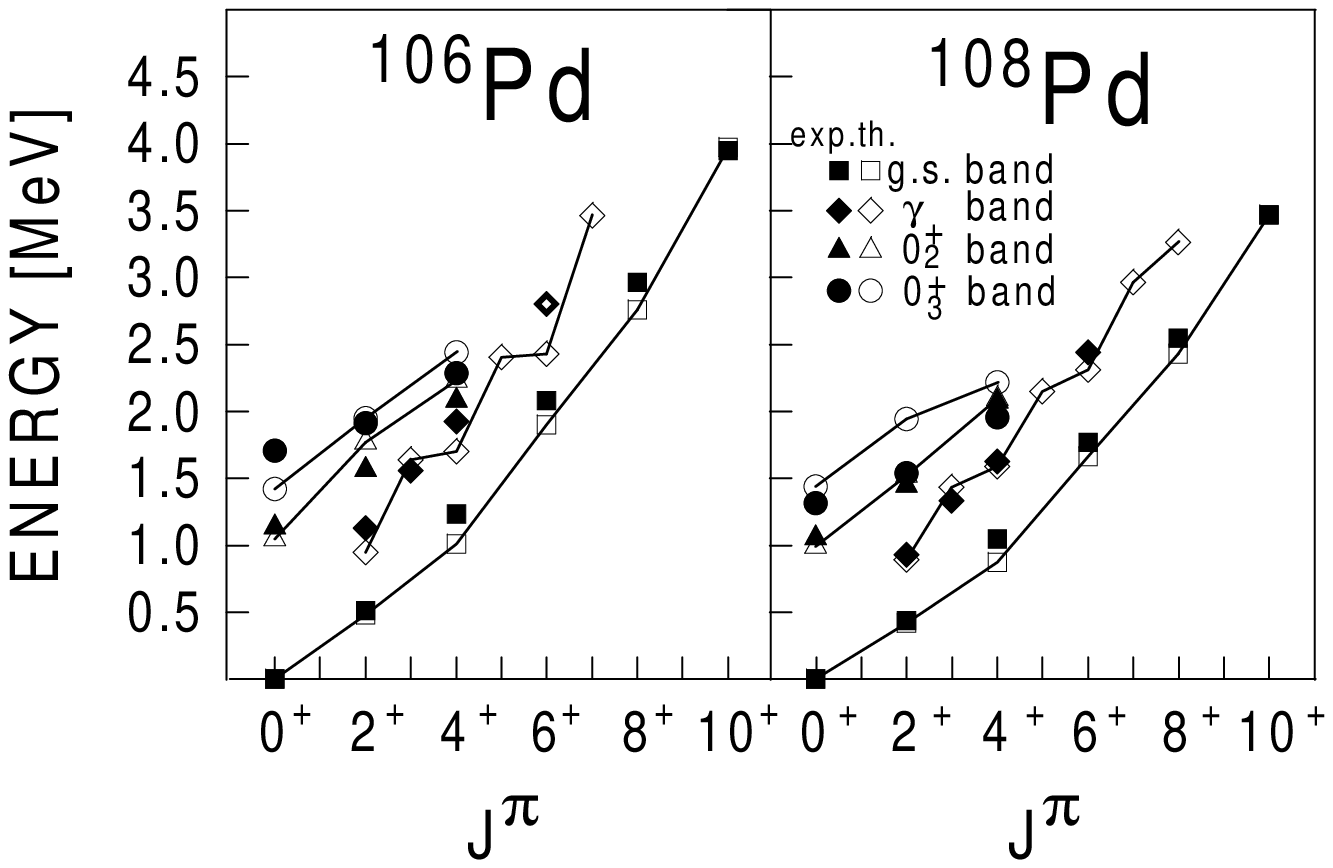}}

\newpage

\framebox{Fig.~13}

\vspace{2cm}
{\includegraphics*{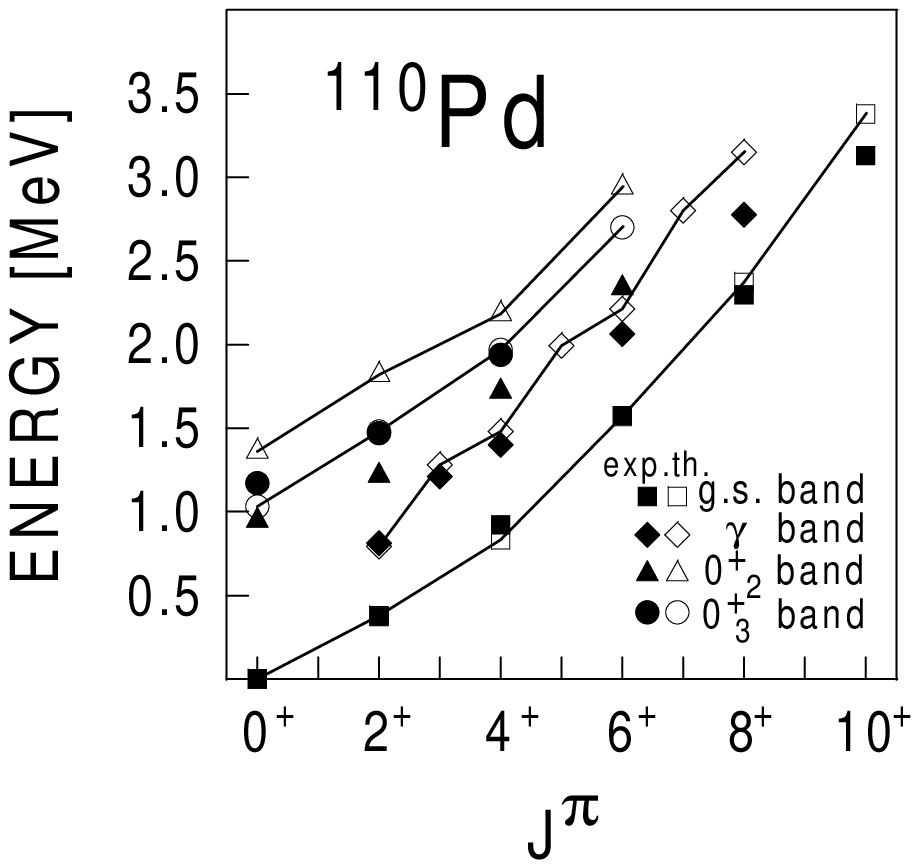}}

\newpage

\framebox{Fig.~14}

\vspace{2cm}
{\includegraphics*{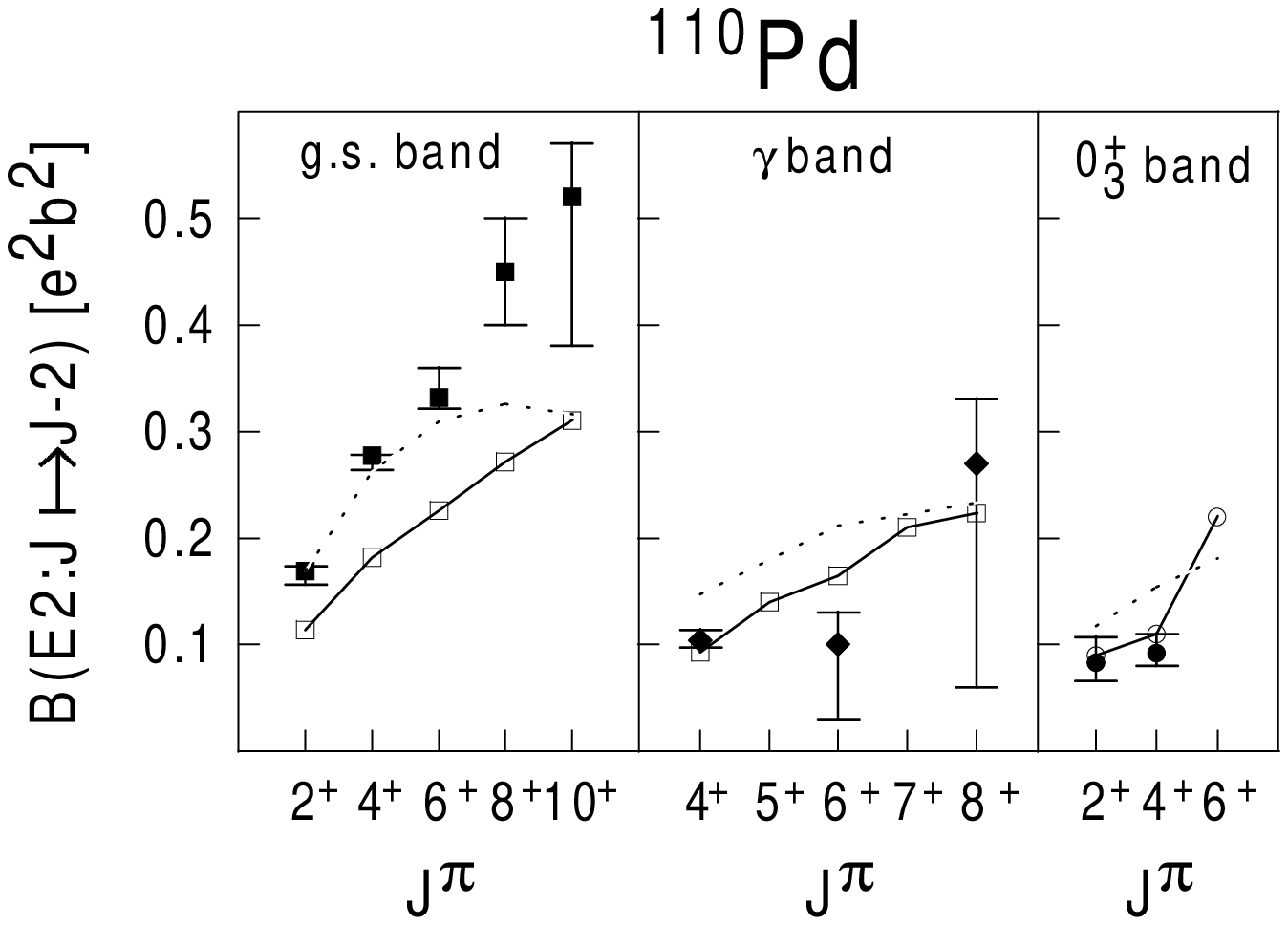}}
\newpage

\framebox{Fig.~15}

\vspace{2cm}
{\includegraphics*{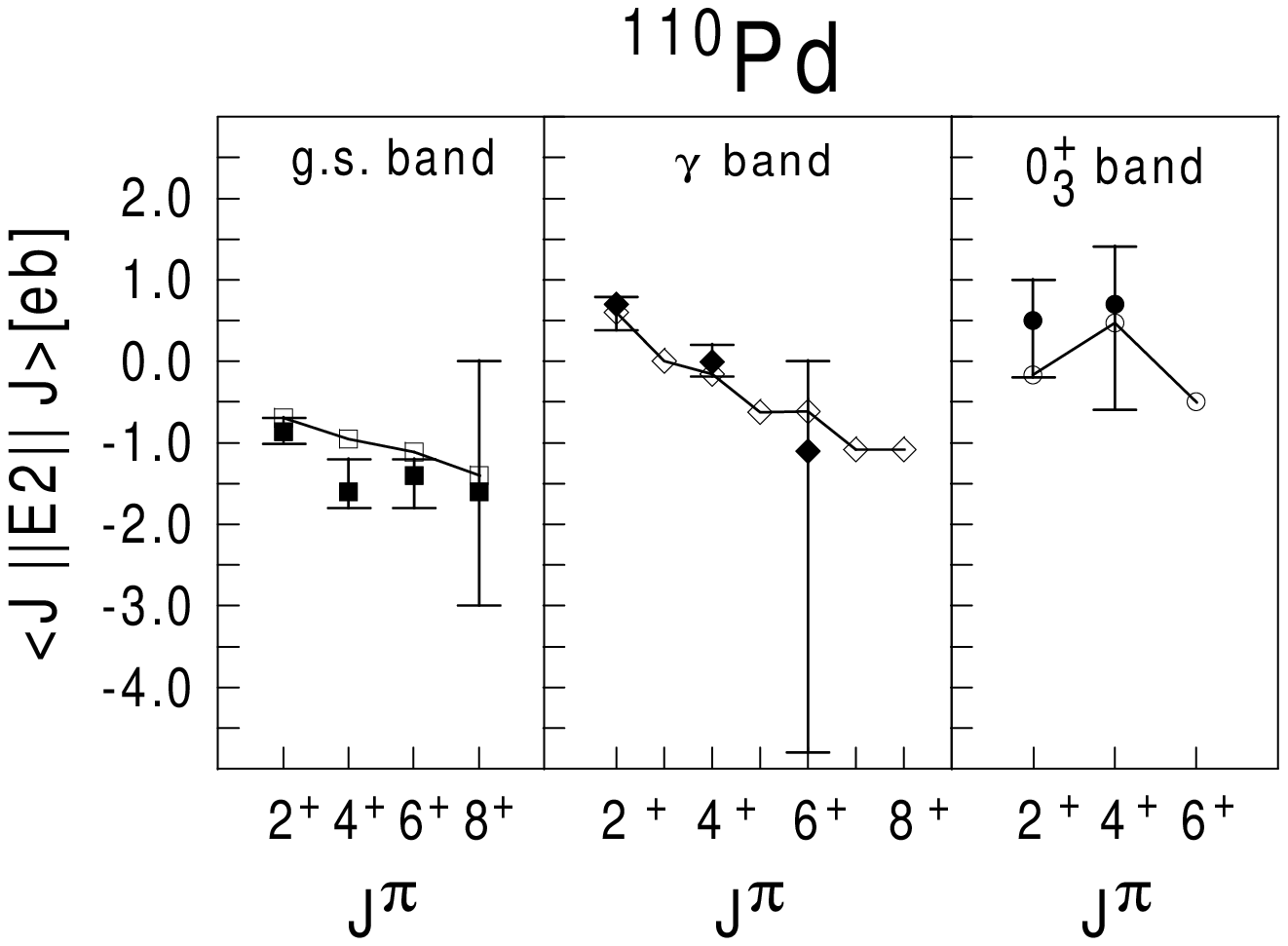}}
\newpage

\framebox{Fig.~16}

\vspace{2cm}
{\includegraphics*{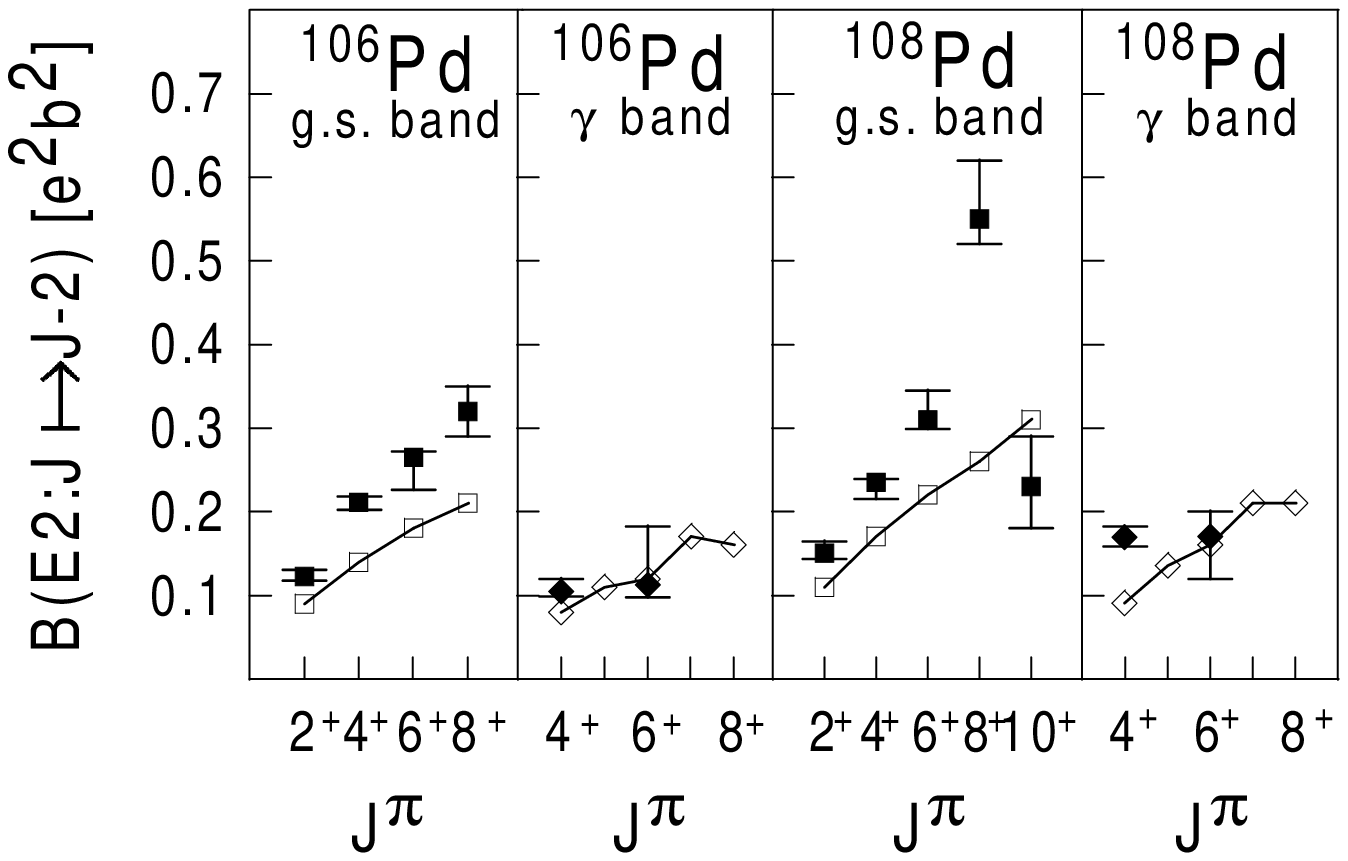}}
\newpage

\framebox{Fig.~17}

\vspace{2cm}
{\includegraphics*{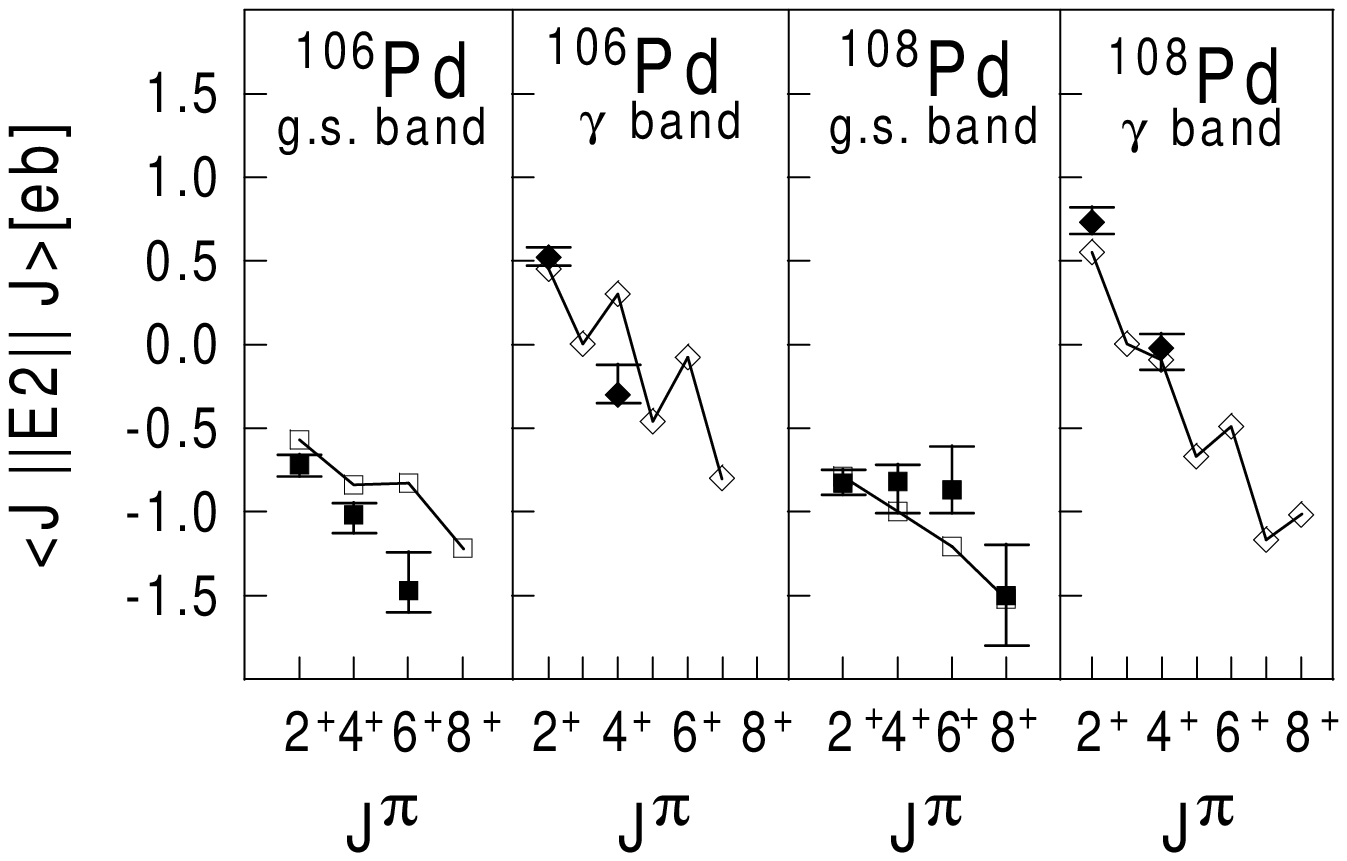}}
\newpage

\framebox{Fig.~18}

\vspace{2cm}
{\includegraphics*{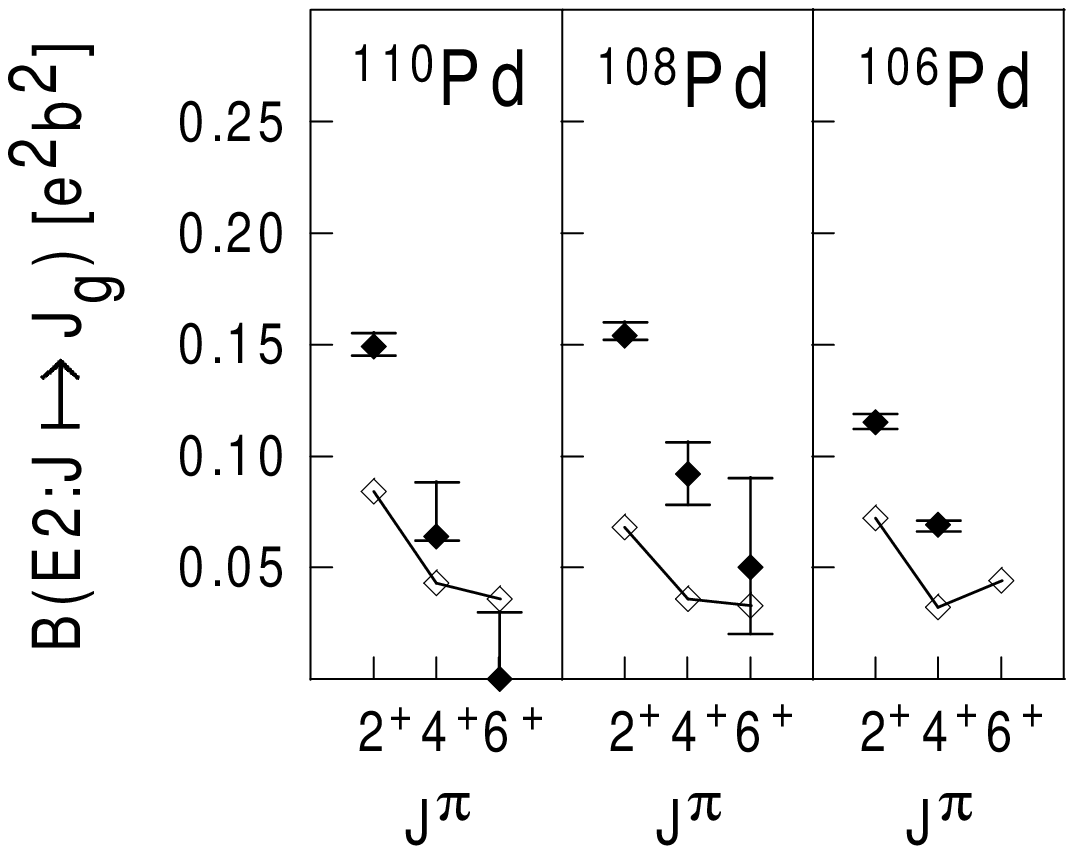}}

\end{document}